\documentclass[conference,10pt]{IEEEtran}
\usepackage{graphicx}
\usepackage{lipsum}
\usepackage{amsmath,bm,amsbsy,amssymb}
\usepackage{mathabx}
\usepackage{fontspec}
\usepackage[dvipsnames]{xcolor}
\usepackage[labelformat=simple]{subcaption}

\usepackage{color}
\usepackage{algorithm}
\usepackage{algpseudocode}
\usepackage{soul}
\soulregister\cite7
\soulregister\ref7
\soulregister\ac7
\soulregister\Ac7
\soulregister\st7
\sethlcolor{Cerulean}
\usepackage{lipsum}
\usepackage[noadjust]{cite}
\usepackage{algorithm,algcompatible,amsmath}
\usepackage{comment}
\usepackage{tikz}
\usepackage{pgfplots}
\pgfplotsset{compat=1.18}

\usepackage{acronym}
\newacro{adc}[ADC]{analog-to-digital converter}
\newacro{adcs}[ADCs]{analog-to-digital converters}
\newacro{dac}[DAC]{digital-to-analog converter}
\newacro{dacs}[DACs]{digital-to-analog converters}
\newacro{snr}[SNR]{signal-to-noise ratio}
\newacro{ber}[BER]{bit error rate}
\newacro{ser}[SER]{symbol error rate}
\newacro{mse}[MSE]{mean squared error}
\newacro{avb}[AVB]{audio video bridging}
\newacro{ofdm}[OFDM]{orthogonal frequency division multiplexing}
\newacro{zpofdm}[ZP-OFDM]{zero-padding OFDM}
\newacro{cpofdm}[CP-OFDM]{cyclic prefix OFDM}
\newacro{dsofdm}[DS-OFDM]{direct sequence OFDM}
\newacro{psk}[PSK]{phase-shift keying}
\newacro{qpsk}[QPSK]{quaternary PSK}
\newacro{bpsk}[BPSK]{binary PSK}
\newacro{qam}[QAM]{quadrature amplitude modulation}
\newacro{fft}[FFT]{fast Fourier transform}
\newacro{dft}[DFT]{discrete Fourier transforms}
\newacro{ls}[LS]{least squares}
\newacro{mrc}[MRC]{maximum ratio combining}
\newacro{dmrc}[DMRC]{differential maximum ratio combining}
\newacro{crc}[CRC]{cyclic redundancy check}
\newacro{spl}[SPL]{sound pressure level}
\newacro{psd}[PSD]{power spectral density}
\newacro{ue}[UE]{user equipment}
\newacro{bs}[BS]{base station}
\newacro{aoa}[AoA]{angles of arrival}
\newacro{act}[ACT]{acoustic communications testbed}
\newacro{sdma}[SDMA]{spatial division multiple access}
\newacro{ici}[ICI]{inter-carrier interference}
\newacro{foc}[FOC]{Frequency Offset Compensation}
\newacro{cdf}[CDF]{cumulative distribution function}
\newacro{dfe}[DFE]{decision-feedback equalizer}
\newacro{isi}[ISI]{inter-symbol interference}
\newacro{pll}[PLL]{phase-locked loop}
\newacro{dll}[DLL]{delay-locked loop}
\newacro{rls}[RLS]{recursive least squares}
\newacro{lms}[LMS]{least mean squares}
\newacro{lms}[LMS]{least mean squares}
\newacro{tr}[TR]{time-reversal}
\newacro{pc}[PC]{phase-conjugate}
\newacro{simo}[SIMO]{single-input multiple-output}
\newacro{mimo}[MIMO]{multiple-input multiple-output}
\newacro{csi}[CSI]{channel state information}
\newacro{idft}[IDFT]{inverse discrete Fourier transform}
\newcommand{\plotsystemblock}{

\tikzset{every picture/.style={line width=0.75pt}} %set default line width to 0.75pt        

\begin{tikzpicture}[x=0.75pt,y=0.75pt,yscale=-1,xscale=1]
\draw   (337,269) -- (372,269) -- (372,309) -- (337,309) -- cycle ;
\draw   (337,389) -- (372,389) -- (372,429) -- (337,429) -- cycle ;
\draw   (182,389) -- (252,389) -- (252,429) -- (182,429) -- cycle ;
\draw   (182,269) -- (252,269) -- (252,309) -- (182,309) -- cycle ;
\draw    (157,289) -- (157,409) ;
\draw    (73,349) -- (95,349) ;
\draw [shift={(98,349)}, rotate = 180] [fill={rgb, 255:red, 0; green, 0; blue, 0 }  ][line width=0.08]  [draw opacity=0] (8.93,-4.29) -- (0,0) -- (8.93,4.29) -- cycle    ;
\draw  [dash pattern={on 0.84pt off 2.51pt}]  (354,339) -- (354,359) ;
\draw [color={rgb, 255:red, 208; green, 2; blue, 27 }  ,draw opacity=1 ][line width=0.75]  [dash pattern={on 4.5pt off 4.5pt}]  (387,227.72) -- (387,437) ;
\draw  [dash pattern={on 0.84pt off 2.51pt}]  (215,337) -- (215,357) ;
\draw   (534,269) -- (569,269) -- (569,429) -- (534,429) -- cycle ;
\draw    (611.5,386.5) -- (611.5,369.5) ;
\draw [shift={(611.5,366.5)}, rotate = 90] [fill={rgb, 255:red, 0; green, 0; blue, 0 }  ][line width=0.08]  [draw opacity=0] (8.93,-4.29) -- (0,0) -- (8.93,4.29) -- cycle    ;
\draw  [dash pattern={on 0.84pt off 2.51pt}]  (442,339) -- (442,359) ;
\draw [color={rgb, 255:red, 208; green, 2; blue, 27 }  ,draw opacity=1 ][line width=0.75]  [dash pattern={on 4.5pt off 4.5pt}]  (513,227.72) -- (513,437) ;
\draw   (594,349) .. controls (594,339.34) and (601.84,331.5) .. (611.5,331.5) .. controls (621.16,331.5) and (629,339.34) .. (629,349) .. controls (629,358.66) and (621.16,366.5) .. (611.5,366.5) .. controls (601.84,366.5) and (594,358.66) .. (594,349) -- cycle ;
\draw   (277,289) .. controls (277,279.34) and (284.84,271.5) .. (294.5,271.5) .. controls (304.16,271.5) and (312,279.34) .. (312,289) .. controls (312,298.66) and (304.16,306.5) .. (294.5,306.5) .. controls (284.84,306.5) and (277,298.66) .. (277,289) -- cycle ;
\draw   (277,409) .. controls (277,399.34) and (284.84,391.5) .. (294.5,391.5) .. controls (304.16,391.5) and (312,399.34) .. (312,409) .. controls (312,418.66) and (304.16,426.5) .. (294.5,426.5) .. controls (284.84,426.5) and (277,418.66) .. (277,409) -- cycle ;
\draw    (294.5,326.5) -- (294.5,309.5) ;
\draw [shift={(294.5,306.5)}, rotate = 90] [fill={rgb, 255:red, 0; green, 0; blue, 0 }  ][line width=0.08]  [draw opacity=0] (8.93,-4.29) -- (0,0) -- (8.93,4.29) -- cycle    ;
\draw    (294.5,446.5) -- (294.5,429.5) ;
\draw [shift={(294.5,426.5)}, rotate = 90] [fill={rgb, 255:red, 0; green, 0; blue, 0 }  ][line width=0.08]  [draw opacity=0] (8.93,-4.29) -- (0,0) -- (8.93,4.29) -- cycle    ;
\draw   (407,269) -- (499,269) -- (499,309) -- (407,309) -- cycle ;
\draw   (407,389) -- (499,389) -- (499,429) -- (407,429) -- cycle ;
\draw   (98,329) -- (133,329) -- (133,369) -- (98,369) -- cycle ;
\draw    (372,289) -- (404,289) ;
\draw [shift={(407,289)}, rotate = 180] [fill={rgb, 255:red, 0; green, 0; blue, 0 }  ][line width=0.08]  [draw opacity=0] (8.93,-4.29) -- (0,0) -- (8.93,4.29) -- cycle    ;
\draw    (372,409) -- (404,409) ;
\draw [shift={(407,409)}, rotate = 180] [fill={rgb, 255:red, 0; green, 0; blue, 0 }  ][line width=0.08]  [draw opacity=0] (8.93,-4.29) -- (0,0) -- (8.93,4.29) -- cycle    ;
\draw    (499,289.22) -- (531,289.22) ;
\draw [shift={(534,289.22)}, rotate = 180] [fill={rgb, 255:red, 0; green, 0; blue, 0 }  ][line width=0.08]  [draw opacity=0] (8.93,-4.29) -- (0,0) -- (8.93,4.29) -- cycle    ;
\draw    (499,409.22) -- (531,409.22) ;
\draw [shift={(534,409.22)}, rotate = 180] [fill={rgb, 255:red, 0; green, 0; blue, 0 }  ][line width=0.08]  [draw opacity=0] (8.93,-4.29) -- (0,0) -- (8.93,4.29) -- cycle    ;
\draw   (654,329) -- (746,329) -- (746,369) -- (654,369) -- cycle ;
\draw   (770,329) -- (862,329) -- (862,369) -- (770,369) -- cycle ;
\draw    (133.08,348.83) -- (155.08,348.83) ;
\draw [shift={(158.08,348.83)}, rotate = 180] [fill={rgb, 255:red, 0; green, 0; blue, 0 }  ][line width=0.08]  [draw opacity=0] (8.93,-4.29) -- (0,0) -- (8.93,4.29) -- cycle    ;
\draw    (157,289) -- (179,289) ;
\draw [shift={(182,289)}, rotate = 180] [fill={rgb, 255:red, 0; green, 0; blue, 0 }  ][line width=0.08]  [draw opacity=0] (8.93,-4.29) -- (0,0) -- (8.93,4.29) -- cycle    ;
\draw    (157,409) -- (179,409) ;
\draw [shift={(182,409)}, rotate = 180] [fill={rgb, 255:red, 0; green, 0; blue, 0 }  ][line width=0.08]  [draw opacity=0] (8.93,-4.29) -- (0,0) -- (8.93,4.29) -- cycle    ;
\draw    (252,289) -- (274,289) ;
\draw [shift={(277,289)}, rotate = 180] [fill={rgb, 255:red, 0; green, 0; blue, 0 }  ][line width=0.08]  [draw opacity=0] (8.93,-4.29) -- (0,0) -- (8.93,4.29) -- cycle    ;
\draw    (252,409) -- (274,409) ;
\draw [shift={(277,409)}, rotate = 180] [fill={rgb, 255:red, 0; green, 0; blue, 0 }  ][line width=0.08]  [draw opacity=0] (8.93,-4.29) -- (0,0) -- (8.93,4.29) -- cycle    ;
\draw    (312,289) -- (334,289) ;
\draw [shift={(337,289)}, rotate = 180] [fill={rgb, 255:red, 0; green, 0; blue, 0 }  ][line width=0.08]  [draw opacity=0] (8.93,-4.29) -- (0,0) -- (8.93,4.29) -- cycle    ;
\draw    (312,409) -- (334,409) ;
\draw [shift={(337,409)}, rotate = 180] [fill={rgb, 255:red, 0; green, 0; blue, 0 }  ][line width=0.08]  [draw opacity=0] (8.93,-4.29) -- (0,0) -- (8.93,4.29) -- cycle    ;
\draw    (569,349) -- (591,349) ;
\draw [shift={(594,349)}, rotate = 180] [fill={rgb, 255:red, 0; green, 0; blue, 0 }  ][line width=0.08]  [draw opacity=0] (8.93,-4.29) -- (0,0) -- (8.93,4.29) -- cycle    ;
\draw    (629,349) -- (651,349) ;
\draw [shift={(654,349)}, rotate = 180] [fill={rgb, 255:red, 0; green, 0; blue, 0 }  ][line width=0.08]  [draw opacity=0] (8.93,-4.29) -- (0,0) -- (8.93,4.29) -- cycle    ;
\draw    (746,349) -- (768,349) ;
\draw [shift={(771,349)}, rotate = 180] [fill={rgb, 255:red, 0; green, 0; blue, 0 }  ][line width=0.08]  [draw opacity=0] (8.93,-4.29) -- (0,0) -- (8.93,4.29) -- cycle    ;
\draw    (862,349) -- (884,349) ;
\draw [shift={(887,349)}, rotate = 180] [fill={rgb, 255:red, 0; green, 0; blue, 0 }  ][line width=0.08]  [draw opacity=0] (8.93,-4.29) -- (0,0) -- (8.93,4.29) -- cycle    ;

\draw (217,289) node [font=\Large]   {$\psi _{0}( t)$};
\draw (57,347.92) node  [font=\Large]  {$d[n]$};
\draw (115.5,349) node  [font=\Large]  {$g( t)$};
\draw (217,409) node  [font=\Large]  {$\psi _{M-1}( t)$};
\draw (216.29,243.33) node  [color={rgb, 255:red, 208; green, 2; blue, 27 }  ,opacity=1 ] [align=left] {beamforming};
\draw (453,289) node  [font=\large]  {$h_{0,\text{dn}}( \tau ,t)$};
\draw (551.5,349) node  [font=\Huge]  {$\Sigma $};
\draw (905,343.92) node [font=\Large]   {$\hat{d}[ n]$};
\draw (613,401) node  [font=\Large]  {$n( t)$};
\draw (453.29,243.33) node  [color={rgb, 255:red, 208; green, 2; blue, 27 }  ,opacity=1 ] [align=left] {channel};
\draw (816,349) node   [align=center] {Equalization};
\draw (453,409) node  [font=\large]  {$h_{M-1,\text{dn}}( \tau ,t)$};
\draw (611.5,349) node  [font=\Large]  {$+$};
\draw (294.5,289) node  [font=\Large]  {$\times $};
\draw (294.5,409) node [font=\Large]   {$\times $};
\draw (295,338) node  [font=\Large]  {$e^{j2\pi f_{c} t}$};
\draw (296,460) node  [font=\Large]  {$e^{j2\pi f_{c} t}$};
\draw (354.5,289) node   [align=left] {$\text{Re} \{ \cdot \}$};
\draw (354.5,409) node   [align=left] {$\text{Re}\{\cdot\}$};
\draw (700,349) node   [align=center] {Downshift $\displaystyle \&$\\Synchronization};

\end{tikzpicture}

}

\newcommand{\plotequalization}{

\tikzset{every picture/.style={line width=0.75pt}} %set default line width to 0.75pt        

\begin{tikzpicture}[x=0.75pt,y=0.75pt,yscale=-1,xscale=1]
\draw    (453.74,104.17) -- (498.74,104.17) ;
\draw [shift={(500.74,104.17)}, rotate = 180] [color={rgb, 255:red, 0; green, 0; blue, 0 }  ][line width=0.75]    (10.93,-3.29) .. controls (6.95,-1.4) and (3.31,-0.3) .. (0,0) .. controls (3.31,0.3) and (6.95,1.4) .. (10.93,3.29)   ;
\draw    (175.74,104.17) -- (233.74,104.17) ;
\draw [shift={(235.74,104.17)}, rotate = 180] [color={rgb, 255:red, 0; green, 0; blue, 0 }  ][line width=0.75]    (10.93,-3.29) .. controls (6.95,-1.4) and (3.31,-0.3) .. (0,0) .. controls (3.31,0.3) and (6.95,1.4) .. (10.93,3.29)   ;
\draw    (121.74,104.17) -- (145.74,104.17) ;
\draw    (145.74,104.17) -- (175.74,84.17) ;
\draw    (145.74,84.17) .. controls (155.03,84.39) and (161.98,96.94) .. (164.84,102.43) ;
\draw [shift={(165.74,104.17)}, rotate = 241.93] [color={rgb, 255:red, 0; green, 0; blue, 0 }  ][line width=0.75]    (7.65,-3.43) .. controls (4.86,-1.61) and (2.31,-0.47) .. (0,0) .. controls (2.31,0.47) and (4.86,1.61) .. (7.65,3.43)   ;
\draw   (341.74,94.17) -- (401.74,94.17) -- (401.74,114.17) -- (341.74,114.17) -- cycle ;
\draw   (433.74,104.17) .. controls (433.74,98.64) and (438.22,94.17) .. (443.74,94.17) .. controls (449.26,94.17) and (453.74,98.64) .. (453.74,104.17) .. controls (453.74,109.69) and (449.26,114.17) .. (443.74,114.17) .. controls (438.22,114.17) and (433.74,109.69) .. (433.74,104.17) -- cycle ;
\draw    (437.47,111.53) -- (451.11,97.17) ;
\draw    (450.92,111.35) -- (436.74,97.35) ;

\draw   (500.74,104.17) .. controls (500.74,98.64) and (505.22,94.17) .. (510.74,94.17) .. controls (516.26,94.17) and (520.74,98.64) .. (520.74,104.17) .. controls (520.74,109.69) and (516.26,114.17) .. (510.74,114.17) .. controls (505.22,114.17) and (500.74,109.69) .. (500.74,104.17) -- cycle ;
\draw    (520.74,104.17) -- (544.74,104.17) ;
\draw [shift={(546.74,104.17)}, rotate = 180] [color={rgb, 255:red, 0; green, 0; blue, 0 }  ][line width=0.75]    (10.93,-3.29) .. controls (6.95,-1.4) and (3.31,-0.3) .. (0,0) .. controls (3.31,0.3) and (6.95,1.4) .. (10.93,3.29)   ;
\draw    (606.74,104.17) -- (652.74,104.17) ;
\draw [shift={(654.74,104.17)}, rotate = 180] [color={rgb, 255:red, 0; green, 0; blue, 0 }  ][line width=0.75]    (10.93,-3.29) .. controls (6.95,-1.4) and (3.31,-0.3) .. (0,0) .. controls (3.31,0.3) and (6.95,1.4) .. (10.93,3.29)   ;
\draw    (625.74,104.17) -- (625.74,154.17) ;
\draw    (609.74,154.17) -- (625.74,154.17) ;
\draw [shift={(607.74,154.17)}, rotate = 0] [color={rgb, 255:red, 0; green, 0; blue, 0 }  ][line width=0.75]    (10.93,-3.29) .. controls (6.95,-1.4) and (3.31,-0.3) .. (0,0) .. controls (3.31,0.3) and (6.95,1.4) .. (10.93,3.29)   ;
\draw    (510.74,154.17) -- (546.74,154.17) ;
\draw    (510.74,116.17) -- (510.74,154.17) ;
\draw [shift={(510.74,114.17)}, rotate = 90] [color={rgb, 255:red, 0; green, 0; blue, 0 }  ][line width=0.75]    (10.93,-3.29) .. controls (6.95,-1.4) and (3.31,-0.3) .. (0,0) .. controls (3.31,0.3) and (6.95,1.4) .. (10.93,3.29)   ;
\draw   (239.25,187.17) -- (647.25,187.17) -- (647.25,207.95) -- (239.25,207.95) -- cycle ;
\draw    (625.74,174.17) -- (625.74,185.17) ;
\draw [shift={(625.74,187.17)}, rotate = 270] [color={rgb, 255:red, 0; green, 0; blue, 0 }  ][line width=0.75]    (10.93,-3.29) .. controls (6.95,-1.4) and (3.31,-0.3) .. (0,0) .. controls (3.31,0.3) and (6.95,1.4) .. (10.93,3.29)   ;
\draw    (575.74,187.17) -- (575.74,166.17) ;
\draw [shift={(575.74,164.17)}, rotate = 90] [color={rgb, 255:red, 0; green, 0; blue, 0 }  ][line width=0.75]    (10.93,-3.29) .. controls (6.95,-1.4) and (3.31,-0.3) .. (0,0) .. controls (3.31,0.3) and (6.95,1.4) .. (10.93,3.29)   ;
\draw    (443.74,187.17) -- (443.74,116.17) ;
\draw [shift={(443.74,114.17)}, rotate = 90] [color={rgb, 255:red, 0; green, 0; blue, 0 }  ][line width=0.75]    (10.93,-3.29) .. controls (6.95,-1.4) and (3.31,-0.3) .. (0,0) .. controls (3.31,0.3) and (6.95,1.4) .. (10.93,3.29)   ;
\draw    (369.74,187.17) -- (370.24,117) ;
\draw [shift={(370.25,115)}, rotate = 90.4] [color={rgb, 255:red, 0; green, 0; blue, 0 }  ][line width=0.75]    (10.93,-3.29) .. controls (6.95,-1.4) and (3.31,-0.3) .. (0,0) .. controls (3.31,0.3) and (6.95,1.4) .. (10.93,3.29)   ;
\draw    (257.74,187.17) -- (258.24,117) ;
\draw [shift={(258.25,115)}, rotate = 90.4] [color={rgb, 255:red, 0; green, 0; blue, 0 }  ][line width=0.75]    (10.93,-3.29) .. controls (6.95,-1.4) and (3.31,-0.3) .. (0,0) .. controls (3.31,0.3) and (6.95,1.4) .. (10.93,3.29)   ;
\draw   (237.74,94.17) -- (281.74,94.17) -- (281.74,114.17) -- (237.74,114.17) -- cycle ;
\draw    (281.74,104.17) -- (339.74,104.17) ;
\draw [shift={(341.74,104.17)}, rotate = 180] [color={rgb, 255:red, 0; green, 0; blue, 0 }  ][line width=0.75]    (10.93,-3.29) .. controls (6.95,-1.4) and (3.31,-0.3) .. (0,0) .. controls (3.31,0.3) and (6.95,1.4) .. (10.93,3.29)   ;
\draw   (546.74,144.17) -- (606.74,144.17) -- (606.74,164.17) -- (546.74,164.17) -- cycle ;
\draw   (546.74,94.17) -- (606.74,94.17) -- (606.74,114.17) -- (546.74,114.17) -- cycle ;
\draw    (625.74,154.17) -- (625.74,160.17) ;
\draw  [fill={rgb, 255:red, 0; green, 0; blue, 0 }  ,fill opacity=1 ] (623.35,104.17) .. controls (623.35,102.91) and (624.37,101.89) .. (625.63,101.89) .. controls (626.89,101.89) and (627.92,102.91) .. (627.92,104.17) .. controls (627.92,105.44) and (626.89,106.46) .. (625.63,106.46) .. controls (624.37,106.46) and (623.35,105.44) .. (623.35,104.17) -- cycle ;
\draw  [fill={rgb, 255:red, 0; green, 0; blue, 0 }  ,fill opacity=1 ] (623.46,154.17) .. controls (623.46,152.91) and (624.48,151.88) .. (625.74,151.88) .. controls (627,151.88) and (628.02,152.91) .. (628.02,154.17) .. controls (628.02,155.43) and (627,156.45) .. (625.74,156.45) .. controls (624.48,156.45) and (623.46,155.43) .. (623.46,154.17) -- cycle ;
\draw    (402.14,104.17) -- (431.44,104.17) ;
\draw [shift={(433.44,104.17)}, rotate = 180] [color={rgb, 255:red, 0; green, 0; blue, 0 }  ][line width=0.75]    (10.93,-3.29) .. controls (6.95,-1.4) and (3.31,-0.3) .. (0,0) .. controls (3.31,0.3) and (6.95,1.4) .. (10.93,3.29)   ;

\draw (522.97,117.37) node    {$-$};
\draw (529.42,87.32) node  [font=\normalsize]  {$\hat{d}[ n]$};
\draw (625.42,86.32) node  [font=\normalsize]  {$\tilde{d}[ n]$};
\draw (475.42,89.32) node  [font=\normalsize]  {$x[ n]$};
\draw (528.17,142.32) node  [font=\normalsize]  {$z[ n]$};
\draw (664,165.32) node  [font=\scriptsize]  {$e[ n] =\tilde{d}[ n] -\hat{d}[ n]$};
\draw (470.56,133.52) node  [font=\small]  {$e^{-j\hat{\varphi }_{0}( nT)}$};
\draw (284.14,144.35) node  [font=\small]  {$\frac{\hat{\varphi }_{0}( nT)}{2\pi f_{c}}$};
\draw (105.89,100.9) node  [font=\footnotesize]  {$v( t)$};
\draw (314.96,89.12) node  [font=\footnotesize]  {$y( nT_{s})$};
\draw (204.38,89) node  [font=\footnotesize]  {$v( nT_{s})$};
\draw (371.74,104.17) node    {$\mathbf{a}[ n]$};
\draw (259.74,104.17) node    {$\mathcal{I}$};
\draw (576.74,154.17) node    {$\mathbf{b}[ n]$};
\draw (576.74,104.17) node  [font=\large] [align=left] {decision};
\draw (423.11,197.56) node  [font=\large] [align=left] {Equalizer | Parameter Update};
\draw (510.74,104.17) node    {$+$};
\draw (152.89,66.9) node  [font=\footnotesize]  {$T_{s} \ =T/2$};

\end{tikzpicture}

}

\newcommand{\plotspacemsecdf}{
\begin{tikzpicture}
\begin{axis}[
    width=10cm,
    height=8cm,
    xlabel={Mean-squared error [dB]},
    ylabel={Estimated CDF},
    grid=both,
    ymin=0,
    ymax=1,
    xmin=-18,
    xmax=5,
    legend style={
        at={(0.5,0.98)},
        anchor=north,
    },
]
\addplot[
    very thick,
    const plot,
] table[
    x index=0,
    y index=1,
    col sep=space
]{figs_tikz/data_tikz/SPACE08_2880157F038_S3_do_multiuser_False_bf_method_BF_mse_user_1_cdf.dat};
\addlegendentry{Day 288 (02:00 UTC)}
\addplot[
    very thick,
    dashed,
    const plot,
] table[
    x index=0,
    y index=1,
    col sep=space
]{figs_tikz/data_tikz/SPACE08_2941154F038_S3_do_multiuser_False_bf_method_BF_mse_user_1_cdf.dat};
\addlegendentry{Day 294 (12:00 UTC)}
\addplot[
    very thick,
    dotted,
    const plot,
] table[
    x index=0,
    y index=1,
    col sep=space
]{figs_tikz/data_tikz/SPACE08_3000754F038_S3_do_multiuser_False_bf_method_BF_mse_user_1_cdf.dat};
\addlegendentry{Day 300 (08:00 UTC)}
\addplot[
    blue,
    solid,
    very thick,
    const plot,
] table[
    x index=0,
    y index=1,
    col sep=space
]{figs_tikz/data_tikz/SPACE08_2880157F038_S3_do_multiuser_False_bf_method_TR_mse_user_1_cdf.dat};
\addplot[
    blue,
    very thick,
    dashed,
    const plot,
] table[
    x index=0,
    y index=1,
    col sep=space
]{figs_tikz/data_tikz/SPACE08_2941154F038_S3_do_multiuser_False_bf_method_TR_mse_user_1_cdf.dat};
\addplot[
    blue,
    very thick,
    dotted,
    const plot,
] table[
    x index=0,
    y index=1,
    col sep=space
]{figs_tikz/data_tikz/SPACE08_3000754F038_S3_do_multiuser_False_bf_method_TR_mse_user_1_cdf.dat};
\end{axis}
\node[
    draw,
    fill=white,
    anchor=south west,
    inner sep=4pt
] at ([xshift=-21.8mm,yshift=-30mm]current axis.north) {
\begin{tabular}{ll}
\raisebox{0.4ex}{\tikz{\draw[black, very thick, solid] (0,0) -- (0.7,0);}} & BF \\
\raisebox{0.4ex}{\tikz{\draw[blue, very thick, solid]  (0,0) -- (0.7,0);}} & TR
\end{tabular}
};
\end{tikzpicture}
}

\newcommand{\plotmacemsecdf}{
\begin{tikzpicture}
\begin{axis}[
    width=10cm,
    height=8cm,
    xlabel={Mean-squared error [dB]},
    ylabel={Estimated CDF},
    grid=both,
    ymin=0,
    ymax=1,
    xmin=-30,
    xmax=5,
    legend style={
        at={(0.02,0.9)},
        anchor=north west,
    },
]
\addplot[
    very thick,
    const plot,
] table[
    x index=0,
    y index=1,
    col sep=space
]{figs_tikz/data_tikz/MACE10_1750155F1811_S5_do_multiuser_False_bf_method_BF_mse_user_1_cdf.dat};
\addlegendentry{(08:11 UTC)}
\addplot[
    very thick,
    dashed,
    const plot,
] table[
    x index=0,
    y index=1,
    col sep=space
]{figs_tikz/data_tikz/MACE10_1750155F1822_S5_do_multiuser_False_bf_method_BF_mse_user_1_cdf.dat};
\addlegendentry{(08:22 UTC)}
\addplot[
    very thick,
    dotted,
    const plot,
] table[
    x index=0,
    y index=1,
    col sep=space
]{figs_tikz/data_tikz/MACE10_1750155F1830_S5_do_multiuser_False_bf_method_BF_mse_user_1_cdf.dat};
\addlegendentry{(08:30 UTC)}
\addplot[
    blue,
    very thick,
    const plot,
] table[
    x index=0,
    y index=1,
    col sep=space
]{figs_tikz/data_tikz/MACE10_1750155F1811_S5_do_multiuser_False_bf_method_TR_mse_user_1_cdf.dat};
\addplot[
    very thick,
    blue,
    dashed,
    const plot,
] table[
    x index=0,
    y index=1,
    col sep=space
]{figs_tikz/data_tikz/MACE10_1750155F1822_S5_do_multiuser_False_bf_method_TR_mse_user_1_cdf.dat};
\addplot[
    blue,
    dotted,
    very thick,
    const plot,
] table[
    x index=0,
    y index=1,
    col sep=space
]{figs_tikz/data_tikz/MACE10_1750155F1830_S5_do_multiuser_False_bf_method_TR_mse_user_1_cdf.dat};
\end{axis}
\node[
    draw,
    fill=white,
    anchor=south west,
    inner sep=4pt
] at ([xshift=2mm,yshift=2mm]current axis.south west) {
\begin{tabular}{ll}
\raisebox{0.4ex}{\tikz{\draw[black, very thick, solid] (0,0) -- (0.7,0);}} & BF \\
\raisebox{0.4ex}{\tikz{\draw[blue, very thick, solid]  (0,0) -- (0.7,0);}} & TR
\end{tabular}
};
\node[
    anchor=south west,
    font=\bfseries
] at ([xshift=5mm,yshift=-6mm]rel axis cs:0.02,0.98)
{Day 176};
\end{tikzpicture}
}

\title{Transmit Beamforming for High-Rate Underwater Acoustic Communications}
\author{\IEEEauthorblockN{Diego A. Cuji$^*$, \textit{Member, IEEE}, Andrew C. Singer$^*$, \textit{Fellow, IEEE}, and Milica Stojanovic$^\dag$, \textit{Fellow, IEEE} }
\IEEEauthorblockA{$^*$ Stony Brook University, Stony Brook, NY USA}
\IEEEauthorblockA{$^\dag$ Northeastern University, Boston, MA USA}}

\begin{document}

% \markboth{Journal of Oceanic Engineering,~Vol.~XX, No.~X, Month~2025}%
\maketitle
\begin{abstract}
    % Transmit beamforming for underwater acoustic communication is challenging because it requires perfect knowledge of the channel to the receiver in advance. In practice, channel estimates must be learned through feedback and are often noisy or outdated because of feedback delay and channel variation. In this paper, we investigate angle-based beamforming strategies for a single-user link that reduce dependence on full channel knowledge by exploiting stable components of the geometric structure in the propagation field. In particular, we focus on scenarios in which there exists a dominant path that remains relatively stable over time, making it a suitable candidate for transmit beamforming. Experimental results using the SPACE and MACE data sets demonstrate the effectiveness of the proposed method in terms of data-detection mean-squared error and bit error rate.
    Transmit beamforming for underwater acoustic communications is challenging because channel state information must be obtained through feedback and may be noisy or outdated. In this paper, we investigate an angle-based beamforming strategy for a single-user link that reduces reliance on full channel knowledge by exploiting stable geometric features of the propagation field. In particular, the beam is steered toward a principal propagation path that remains relatively stable over time. Experimental results using the SPACE and MACE data sets demonstrate reliable communication with excellent data-detection mean-squared error and zero bit errors.
\end{abstract}
\vspace{-1.4em}
\section{\label{sec:1} Introduction}
Underwater acoustic communication over a single-user link remains challenging because of the limited bandwidth and time variation of the channel. Multipath propagation creates extended delay spread and severe intersymbol interference, while platform motion and environmental variability introduce Doppler distortions. These effects make transmit beamforming particularly challenging in phase-coherent systems.

Previous work has primarily relied on reciprocity-based transmit beamforming using \ac{tr} mirrors or phase-conjugate arrays \cite{parvulescu1965reproducibilityocean,parvulescu1995oceanacoustics,fink1997timereversal,fink2000timereversal}. These techniques were first developed for \ac{simo} communications \cite{kuperman1998phaseconjugation,rouseff2001phaseconjugation,edelmann2002initialtimereversal,edelman2005timereversal,stojanovic2005retrofocusing,song2006spatialdiversitytimereversal} and later extended to \ac{mimo} and multi-user systems \cite{song2006mimotimereversal,song2007multiusertimereversal,song2010crosstalk,song2014crosstalkleastsquares,shimura2015multiusertimereversal,song2016overviewtimereversal,shimura2021highratemimo,kida2023mimo}. However, these methods rely on \ac{csi} obtained through feedback, which may become inaccurate because of channel time variation.

% To address these challenges, prior work on underwater uplink (mobile to base) and downlink (base to mobile) transmit beamforming has primarily focused on reciprocity, leveraging \ac{tr} mirrors or phase-conjugate arrays \cite{fink1997timereversal,fink2000timereversal}. The concept of \ac{tr} involves retransmitting a received signal in time-reversed order, or equivalently in phase-conjugated form in the frequency domain. Its application to underwater acoustic systems was first developed for \ac{simo} communications \cite{kuperman1998phaseconjugation,rouseff2001phaseconjugation,edelmann2002initialtimereversal,edelman2005timereversal,stojanovic2005retrofocusing,song2006spatialdiversitytimereversal}, and later extended to \ac{mimo} and multi-user systems \cite{song2006mimotimereversal,song2007multiusertimereversal,song2010crosstalk,song2014crosstalkleastsquares,shimura2015multiusertimereversal,song2016overviewtimereversal,shimura2021highratemimo,kida2023mimo}. Although \ac{tr} methods have shown promising performance, they rely on accurate channel state information, which in practice is obtained through feedback and can therefore be noisy and outdated in time-varying channels.

In this article, we investigate an angle-based beamforming method for single-carrier underwater acoustic communications. The framework exploits the stable components of the propagation field by steering a beam toward the principal propagation path. Unlike our previous work, which developed angle-based transmit beamforming for OFDM systems \cite{cuji2023bftx}, the present paper extends the framework to single-carrier communications by integrating transmit beamforming with front-end synchronization, Doppler compensation, and equalization. The proposed method is validated using the SPACE and MACE datasets, demonstrating reliable performance for both single-user and asynchronous multi-user communication scenarios.

% In this article, we investigate an angle-based transmit beamforming method for single-user, single-carrier underwater acoustic communications. The method exploits stable elements of the geometric structure in the propagation field to reduce reliance on full channel state information. The strategy consists of steering a beam toward the angle of the principal path, while the receiver performs front-end synchronization, Doppler compensation, and adaptive equalization. Experimental results using underwater acoustic data demonstrate that the proposed approach can support reliable high-rate downlink communication.

The rest of the article is organized as follows. In Sec.~\ref{sec:signal_model}, we describe the signal, channel, and system models. We present the experimental results in Sec.~\ref{sec:experimental}, and the conclusions are summarized in Sec.~\ref{sec:conclusion}.

\section{Signal, Channel, and System Models\label{sec:signal_model}}
We consider a base station equipped with a vertical linear array and a single user with a single element. The array consists of $M$ elements equally spaced by $\delta$. At the base station, data symbols $d[n]$ are pulse-shaped by $g(t)$, e.g., a raised cosine filter with roll-off factor $\alpha_{rc}$, and symbol period $T=1/R$, where $R$ is the symbol rate. Each transmitting element applies an additional beamforming filter $\psi_m(t)$, resulting in the effective transmit pulse\footnote{The symbol $\star$ represents convolution.}
% \vspace{-0.5em}
\begin{equation}
    g_m(t) = g(t) \star \psi_m(t) \triangleq \int_{-\infty}^\infty g(\xi) \psi_m(t-\xi) d\xi
\end{equation}
used to form the $m$-th transmitted baseband signal
% \vspace{-0.5em}
\begin{equation}
    u_m(t) = \sum_n d[n]\, g_m(t-nT), \quad m=0,\ldots,M-1
\end{equation}
and the passband signals $s_m(t)=\mathrm{Re}\left\{u_m(t)e^{j2\pi f_c t}\right\}$, where $f_c$ is the center frequency. On the user's end, the received passband signal is modeled as
\begin{equation}
    r(t) = \bar{r}(t) + n(t)
    \label{eq:r_m_t}
\end{equation}
where $n(t)$ represents the additive noise and
% \vspace{-0.5em}
\begin{equation}
    \bar{r}(t) = \sum_{m=0}^{M-1} \int_{-\infty}^\infty h_{m,\mathrm{dn}}(\xi,t) s_m(t-\xi) d\xi
\end{equation}
where $h_{m,\mathrm{dn}}(\tau,t)$ is the time-varying impulse response observed on the downlink (from the $m$-th transmitting element to the user)
\begin{equation}
    h_{m,\mathrm{dn}}(\tau,t) = \sum_{p=0}^{P-1} h_{p,\mathrm{dn}}^m(t) \delta\left(\tau - \tau_{p,\mathrm{dn}}^m(t)\right)
\end{equation}
where the terms $h_{p,\mathrm{dn}}^m(t)$, $\tau_{p,\mathrm{dn}}^m(t)$, and $P$ are the path gains, path delays, and number of propagation paths, respectively. After coherent down-conversion and coarse time synchronization, the received equivalent baseband signal can be expressed as
% \vspace{-0.5em}
\begin{equation}
    \begin{aligned}
        v(t) = \sum_n & d[n] h_{\mathrm{eq}}(t-nT,t) + w(t)
    \end{aligned}
    \label{eq:v_m_t}
\end{equation}
\vspace{-0.5em}
where $w(t)$ is the additive complex-baseband noise and
\begin{equation}
    \begin{aligned}
        h_{\mathrm{eq}}(\tau,t) &= \sum_{m=0}^{M-1} \int_{-\infty}^\infty \underline{h}_{m,\mathrm{dn}}(\xi,t) g_m(\tau-\xi) d\xi
    \end{aligned}
\end{equation}
is the equivalent composite baseband channel response, where
\vspace{-1em}
\begin{equation}
    \underline{h}_{m,\mathrm{dn}}(\tau,t) = \sum_{p=0}^{P-1} h_{p,\mathrm{dn}}^m e^{-j2\pi f_c \tau_{p,\mathrm{dn}}^m(t)} \delta\left(\tau-\tau_{p,\mathrm{dn}}^m(t)\right) 
\end{equation}
represents the equivalent baseband channel impulse response. The path gains are approximated as $ h_{p,\mathrm{dn}}^m(t) \approx h_{p,\mathrm{dn}}^m$. Under the far-field plane-wave assumption, the static delay of the $p$-th propagation path across the array satisfies
$ \tau_{p,\mathrm{dn}}^m = \tau_{p,\mathrm{dn}}^0 + m\Delta\tau_p$, where $ \Delta\tau_p = \frac{\delta}{c}\sin\theta_p$ is the incremental delay associated with the angle of arrival $\theta_p$, and $c$ denotes the speed of sound in water. The time-varying path delays are modeled as $ \tau_{p,\mathrm{dn}}^m(t) = \tau_{p,\mathrm{dn}}^m + \epsilon_{p,\mathrm{dn}}(t)$, where $\epsilon_{p,\mathrm{dn}}(t)$ denotes the Doppler-induced path-specific delay variation. The equivalent composite baseband channel response can be re-written as
\begin{equation}
    h_\mathrm{eq}(\tau,t) = \sum_{m=0}^{M-1}\sum_{p=0}^{P-1} c_{p,\mathrm{dn}}^m e^{-j2\pi f_c \epsilon_{p,\mathrm{dn}}(t)} g_m\left(\tau-\tau_{p,\mathrm{dn}}^m(t)\right) 
\end{equation}
where $c_{p,\mathrm{dn}}^m = c_{p,\mathrm{dn}}^0 e^{-j2\pi f_c m \Delta\tau_p}$ and $c_{p,\mathrm{dn}}^0 = h_{p,\mathrm{dn}} e^{-j2\pi f_c \tau_{p,\mathrm{dn}}^0}$ are the equivalent baseband path gains, assuming $h_{p,\mathrm{dn}}^m \approx h_{p,\mathrm{dn}}$.

% where the path gains are approximated as $h_{p,\mathrm{dn}}^m(t) \approx h_{p,\mathrm{dn}}^m$. The path delays can be expressed as $\tau_{p,\mathrm{dn}}^m(t) = \tau_{p,\mathrm{dn}}^m + \epsilon_{p,\mathrm{dn}}(t)$, where $\epsilon_{p,\mathrm{dn}}(t)$ represents the path-specific Doppler-induced delay variation, and assuming plane-wave propagation $\tau_{p,\mathrm{dn}}^m = \tau_{p,\mathrm{dn}}^0 + m \Delta\tau_p$, where $\Delta\tau_p = \tfrac{\delta}{c}\sin\theta_p$ is the incremental delay across the array and $c$~m/s denotes the speed of sound.

% where $c_{p,\mathrm{dn}}^m = c_{p,\mathrm{dn}} e^{-j2\pi f_c m \Delta\tau_p}$ and $c_{p,\mathrm{dn}} = h_{p,\mathrm{dn}} e^{-j2\pi f_c \tau_{p,\mathrm{dn}}^0}$ is the equivalent baseband path gain. Note that if the channel is considered time-invariant, then the received signal reduces to $ v(t) = \sum_n d[n] h_\mathrm{eq}(t-nT) + w(t)$, where $h_\mathrm{eq}(t) = \sum_{m=0}^{M-1}\sum_{p=0}^{P-1} c_{p,\mathrm{dn}}^m g_m(t-\tau_{p,\mathrm{dn}}^m)$.
% h_{p,\mathrm{dn}}^m e^{-j2\pi f_c \tau_{p,\mathrm{dn}}^m}
\subsection{Transmit Beamforming\label{sec:algorithm}}

Prior to downlink transmission, the user sends an uplink pilot signal to the transmitter array to estimate the uplink \ac{csi}. Conventional reciprocity-based beamforming computes the transmit beamformer directly from the uplink \ac{csi}, making it sensitive to uplink/downlink mismatch caused by channel time variation. Instead, the proposed method estimates only the angle of the principal propagation path from the uplink channel estimate and uses this angle to compute the transmit beamformer. The angle estimation procedure follows \cite{cuji2023bftx} and only requires an uplink pilot longer than the channel multipath spread. Once the principal path angle $\theta_0$ is identified, its corresponding incremental delay is computed as

% We focus our work on two beamforming strategies: channel-based beamforming and angle-based beamforming. The first one includes: optimal beamforming and time-reversal. The second one includes: beamforming towards the principal path angle and null-steering.

% \subsection{Optimal Beamforming}

% Optimal transmit beamforming maximizes the \ac{snr} under a transmit power constraint. The solution is given by

% \begin{equation}
%     \phi_m(f) \triangleq \alpha(f) H_{m,\mathrm{dn}}^*(f)
% \end{equation}
% where $\alpha(f)$ is a normalization constant such that
% \begin{equation}
%     \sum_{m=0}^{M-1} \int |\phi_m(f)|^2 df = P_{tx}
% \end{equation}
% The time-domain beamforming filter is computed using the inverse Fourier transform $\varphi_m(t) = \mathcal{F}^{-1} \left\lbrace \phi_m(f) \right\rbrace$. Note that the optimal beamforming frequency coefficients require knowing the downlink channel, which in practical situations is not available.

% \subsection{Time-reversal}

% The time-reversal or phase-conjugation techniques consist of computing the beamforming filters as
% \begin{equation}
%     \phi_m(f) \triangleq \alpha_{\mathrm{TR}}(f)H_{m,\mathrm{up}}^*(f)
%     % \varphi_m(t) = h_{m,\mathrm{up}}^*(- t)
% \end{equation}
% where $H_{m,\mathrm{up}}(f)$ is the uplink transfer function between the receiver and the $m$-th transmitting element. In practice, the true uplink responses have to be estimated.

% % \subsection{Angle-based Beamforming}
% \subsection{Principal path's angle beamforming}

\begin{equation}
    \Delta\tau_0 = \frac{\delta}{c}\sin\theta_0
\end{equation}
The beamforming vector in the frequency domain, $\Phi(f) = \begin{bmatrix}
    \phi_0(f) & \phi_1(f) & \ldots & \phi_{M-1}(f)
\end{bmatrix}^\top$, is computed as
\begin{equation}
    \begin{aligned}
        \Phi(f) &\triangleq \frac{1}{\sqrt{M}}\mathbf{s}_M\left(2\pi f \Delta\tau_0\right)
    \end{aligned}
    \label{eq:phi_f_dn}
\end{equation}
where $\mathbf{s}_M(\chi) = \begin{bmatrix}
    1 & e^{-j\chi} & \ldots & e^{-j(M-1)\chi}
\end{bmatrix}^\top$ is the steering vector. Defining \eqref{eq:phi_f_dn} ensures unit norm transmit power\footnote{$(\cdot)^\prime$ and $(\cdot)^*$ denote conjugate transpose and complex conjugate, respectively.} $\Phi^\prime(f) \Phi(f) = 1$.

\begin{figure}[ht]
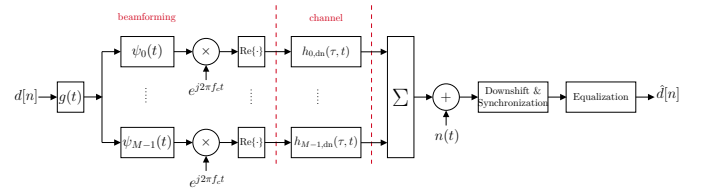

    \centering
    % \scalebox{0.75}{\plotsystemblock}
    \scalebox{0.38}{\plotsystemblock}
    % \caption{System block diagram. The data symbols $d[k]$ are pulse-shaped by $g(t)$ and then pass through $M$ parallel filters $\psi_m(t)$, $m=0,\ldots,M-1$ to produce the baseband signals $u_m(t)$, which are upshifted by $f_c$ resulting in the passband signals $s_m(t)$. The passband signals pass through the time-varying multipath channels $h_{m,\mathrm{dn}}(\tau,t)$. After the channels, the composite signal is added noise $n(t)$ to produce $r(t)$, which is downshifted, time-synchronized, and equalized to produce the estimated data symbols $\hat{d}[k]$.}
    \caption{System block diagram. The data symbols $d[n]$ are pulse-shaped, beamformed, and upconverted for transmission through the channels $h_{m,\mathrm{dn}}(\tau,t)$. The received signal $r(t)$ is downshifted, synchronized, and equalized to obtain the symbol estimates $\hat{d}[n]$.}
    \label{fig:system_block}
\end{figure}

The beamforming filters $\psi_m(t)$ are obtained via the inverse Fourier transform of $\phi_m^*(f)$. In practice, these filters are sampled at $f_s = 1/T_s=N_s/T$, where $N_s\geq 2$ is an integer number of samples per symbol. The beamforming vector is then evaluated at frequencies $f_\ell$, defined as
\begin{equation}
    f_\ell = \begin{cases}
    f_c + \ell \Delta f, & 0 \leq \ell \leq \frac{L}{2} \\
    f_c + (\ell - L) \Delta f, & \frac{L}{2} + 1 \leq \ell \leq L-1
    \end{cases}
\end{equation}
where $L$ is the total number of frequency bins and $\Delta f = (L T_s)^{-1}$ is the frequency bin spacing. Since the transmitted signal occupies only a limited
bandwidth, we define the discrete frequency domain beamforming vector $\Phi_\ell = \Phi(f_\ell)$ as $\Phi_\ell  = \begin{bmatrix}
    \phi_0[\ell] & \phi_1[\ell] & \ldots & \phi_{M-1}[\ell]
\end{bmatrix}^\top$, where $\phi_m[\ell] = \phi_m(f_\ell)$, and set
\[
\Phi_\ell=
\begin{cases}
\dfrac{1}{\sqrt M}\mathbf{s}_M(2\pi f_\ell\Delta\tau_0), & \ell\in\mathcal L_B,\\[1mm]
\mathbf 0, & \ell\notin\mathcal L_B.
\end{cases}
\]
The subset of nonzero weights is
\begin{equation}
\mathcal{L}_B=\left\lbrace
\ell \;\middle|\;
0 \le \ell \le \bar{L}
\;\;\text{or}\;\;
L-\bar{L} \le \ell \le L-1
\right\rbrace,
\end{equation}
where $\bar{L}=\left\lceil\frac{L(1+\alpha_{rc})}{2N_s}\right\rceil$ is the number of frequency bins required to cover one half of the occupied band. The FIR beamforming filters $\psi_m[n] = \psi_m(nT_s)$ are computed via the \ac{idft}. Before applying the \ac{idft}, we obtain $\tilde{\phi}_{m}[\ell] = (-1)^\ell \phi_{m}^*[\ell]$, where the factor $(-1)^\ell$ introduces an $L/2$-sample circular shift of the resulting impulse response. The synthesized beamforming filter for the $m$-th array element is given by
\begin{equation}
    \psi_m[n] = \frac{1}{L}\sum_{\ell=0}^{L-1} \tilde{\phi}_m[\ell] e^{j2\pi \ell n/L} = \frac{1}{L}\sum_{\ell\in \mathcal{L}_B} \tilde{\phi}_m[\ell] e^{j2\pi \ell n/L}.
    \label{eq:phi_time_ifft}
\end{equation}
The overall system model, including transmit beamforming, propagation through the channel, and receiver-side processing, is illustrated in Fig.~\ref{fig:system_block}.
% which are the samples of $\psi_m(t)$, and it is a causal approximation after truncation and delay. In practice, an additional phase-alignment across $\ell$ may be applied to ensure smoothness of $\phi_m[\ell]$ before \eqref{eq:phi_time_ifft}, since each steering vector is defined up to an arbitrary complex phase.

% \subsection{Eigen-beamforming}

% In eigen-beamforming, I observed that without Doppler, the maximum eigenvalue per frequency bin dominates respect to the others.

% However, when there is Doppler it is less dominant.

% I have observed:
% \begin{itemize}
%     \item No Doppler scenarios cause that one eigenvalue to be dominant respect to the others. This is a sign of high coherence.
%     \item Doppler causes loss of coherence (I have looked into papers but have not found any study on this). In other words, the eigenvalue distribution is more spread, i.e., there is no single dominant eigenvalue. MACE shows this as well.
%     \item If the Doppler is specific per path, then that causes a bigger spread in the eigenvalue distribution. If Dopplers are the same, there is still a single dominant eigenvalue.
% \end{itemize}
\vspace{-1.5em}
\subsection{Equalization}
On the user's side, after time-synchronization, the received signal can be written as
\begin{equation}
    \begin{aligned}
        v(t) &= \sum_n d[n] \tilde{g}_0(t - nT - \epsilon_{0,\mathrm{dn}}(t)) e^{-j2\pi f_c \epsilon_{0,\mathrm{dn}}(t)} \\ &\quad+ I(t) +w(t)
        \label{eq:v_t_dn}
    \end{aligned}
\end{equation}
where $w(t)$ is additive noise and
\begin{equation}
    \begin{aligned}
        I(t) &=  \sum_n d[n] \sum_{p\neq 0} \tilde{g}_{p}(t-nT -\delta\tau_{p,\mathrm{dn}}^0- \epsilon_{p,\mathrm{dn}}(t))  e^{-j2\pi f_c \epsilon_{p,\mathrm{dn}}(t)}
    \end{aligned}
\end{equation}
represents an interference term to be mitigated by equalization, with $\delta\tau_{p,\mathrm{dn}}^0 = \tau_{p,\mathrm{dn}}^0 - \tau_{0,\mathrm{dn}}^0$ and
\begin{equation}
    \begin{aligned}
        \tilde{g}_p(t) &= \frac{c_{p,\mathrm{dn}}^0}{\sqrt{M}} \sum_{m=0}^{M-1}  e^{-j2\pi f_c m(\Delta\tau_p - \Delta\tau_0) } g(t - m (\Delta\tau_p - \Delta\tau_0)).
    \end{aligned}
\end{equation}
Note that $\tilde{g}_{0}(t) = \sqrt{M} c_{0,\mathrm{dn}}^0 g(t)$.

We employ a single-channel fractionally-spaced \ac{dfe} with spacing $T_s=T/2$ (two samples per symbol) aided by adaptive delay-tracking \cite{cuji2026bfdfe}.
During the $n$-th symbol interval, two new adaptive resampled (obtained via linear interpolation) input samples are collected
\begin{equation}
    y\!\left(nT+i\frac{T}{2}\right)=
    \mathcal{I}\!\left\{
        v\!\left(nT+i\frac{T}{2}-\frac{\hat{\varphi}_0(nT)}{2\pi f_c}\right)
    \right\}
    \label{eq:y_single_lr}
\end{equation}
for $i=N_1,N_1-1$, where $\hat{\varphi}_0(nT)$ is an estimate of the residual carrier phase of the principal path after resampling. For example, if $\epsilon_{0,\mathrm{dn}}(t)=a_{0,\mathrm{dn}} \cdot t$ and the received signal is resampled as $v\left(\tfrac{t}{1-a_{0,\mathrm{dn}}}\right)$, then $\varphi_0(t) = -\tfrac{2\pi f_c a_{0,\mathrm{dn}}t}{1-a_{0,\mathrm{dn}}}$. Thus, $\tfrac{-\hat{\varphi}_0(nT)}{2\pi f_c}$ represents the corresponding residual delay correction used for adaptive resampling, and implemented through linear interpolation
\begin{equation}
    \mathcal{I} \left\lbrace v\left(t[n]\right) \right\rbrace = \left(1 - \alpha[n]\right) v\left(t_L[n]\right) + \alpha[n] v\left(t_R[n]\right)
    \label{eq:linear_interpolation}
\end{equation}
where $t_L[n] = \lfloor \frac{t[n]}{T_s} \rfloor T_s$ and $t_R[n] = \lceil \frac{t[n]}{T_s} \rceil T_s$ with $\alpha[n] = \frac{t[n] - t_L[n]}{T_s}$.

We then form the feedforward input vector of length $N_f$ as
\begin{equation}
    \mathbf{y}[n]=
    \begin{bmatrix}
        y\!\left(nT+N_1\frac{T}{2}\right) \\
        y\!\left(nT+N_1\frac{T}{2}-\frac{T}{2}\right) \\
        \mathbf{y}[n-1]_{1:N_f-2}
    \end{bmatrix}
    \label{eq:y_single}
\end{equation}
where $\mathbf{y}[n-1]_{1:N_f-2}$ denotes the first $N_f-2$ entries of $\mathbf{y}[n-1]$.

\begin{figure}[ht]
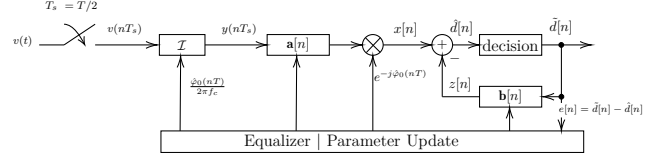

    \centering
    \scalebox{0.52}{\plotequalization}
    \caption{Block diagram of the decision-feedback equalization aided by adaptive delay-tracking on the user's side.}
    \label{fig:equalization}
\end{figure}

Let $\mathbf{a}[n]\in\mathbb{C}^{N_f\times 1}$ and $\mathbf{b}[n]\in\mathbb{C}^{N_b\times 1}$ denote the
feedforward and feedback coefficient vectors, respectively. The feedforward filter output is $x[n]=\mathbf{a}^\prime[n]\mathbf{y}[n] e^{-j\hat{\varphi}_0(nT)}$ and the feedback section operates on the decision vector $\tilde{\mathbf{d}}[n]=
    \begin{bmatrix}
        \tilde{d}[n-1] & \tilde{d}[n-2] & \cdots & \tilde{d}[n-N_b]
    \end{bmatrix}^\top$ to produce $z[n]=\mathbf{b}^\prime[n]\tilde{\mathbf{d}}[n]$. The equalized symbol estimate is then obtained as
\begin{equation}
    \hat{d}[n]=x[n]-z[n].
    \label{eq:dhat_single}
\end{equation}
The decision $\tilde{d}[n]=\mathrm{decision}\{\hat{d}[n]\}$ is the nearest constellation point. The coefficients $\mathbf{c}[n] = \begin{bmatrix}
    \mathbf{a}^\top[n] & -\mathbf{b}^\top[n]
\end{bmatrix}^\top$ are updated recursively as
\begin{equation}
    \mathbf{c}[n+1]=\mathbf{c}[n]+\mathcal{A}\{\mathbf{u}[n],e[n]\}
    \label{eq:c_update_single}
\end{equation}
where $\mathbf{u}[n]\triangleq
    \begin{bmatrix}
        \mathbf{y}^\top[n]e^{-j\hat{\varphi}_0(nT)} & 
        \tilde{\mathbf{d}}^\top[n]
    \end{bmatrix}^\top$, the error is $e[n]=d[n]-\hat{d}[n]$ during training, and $e[n]=\tilde{d}[n]-\hat{d}[n]$ in decision-directed mode, and the term $\mathcal{A}\{\cdot\}$ denotes an adaptive algorithm, e.g., \ac{lms} or \ac{rls}. The phase $\hat{\varphi}_0(nT)$ is tracked using a second order \ac{pll} with constants $K_{f_1}$ and $K_{f_2}$. The block diagram of the equalization process is shown in Fig.~\ref{fig:equalization}.
% \vspace{-2.0em}
\section{Experimental Results\label{sec:experimental}}
\begin{figure}[ht]
    \centering
    \begin{subfigure}[b]{0.24\textwidth}
        \centering
        \includegraphics[width=0.8\linewidth]{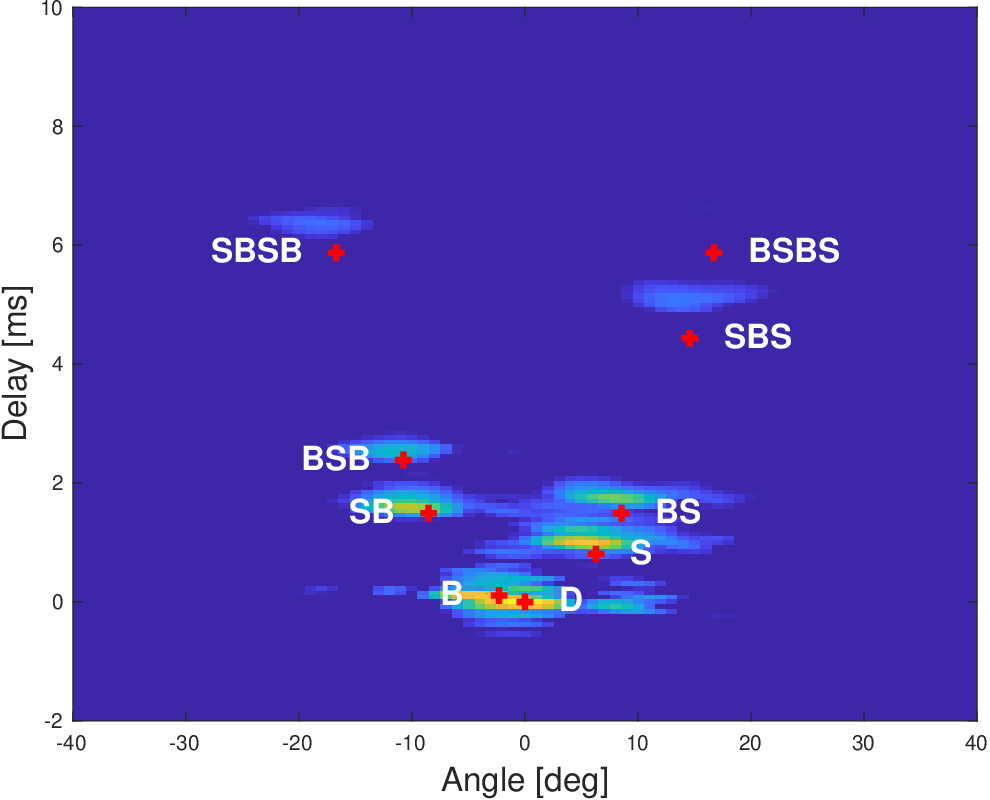}
        \caption{Delay-angle plot}
        \label{fig:SPACE08_average_delay_angle}
    \end{subfigure}
    \begin{subfigure}[b]{0.2\textwidth}
        \centering
        \includegraphics[width=0.8\textwidth]{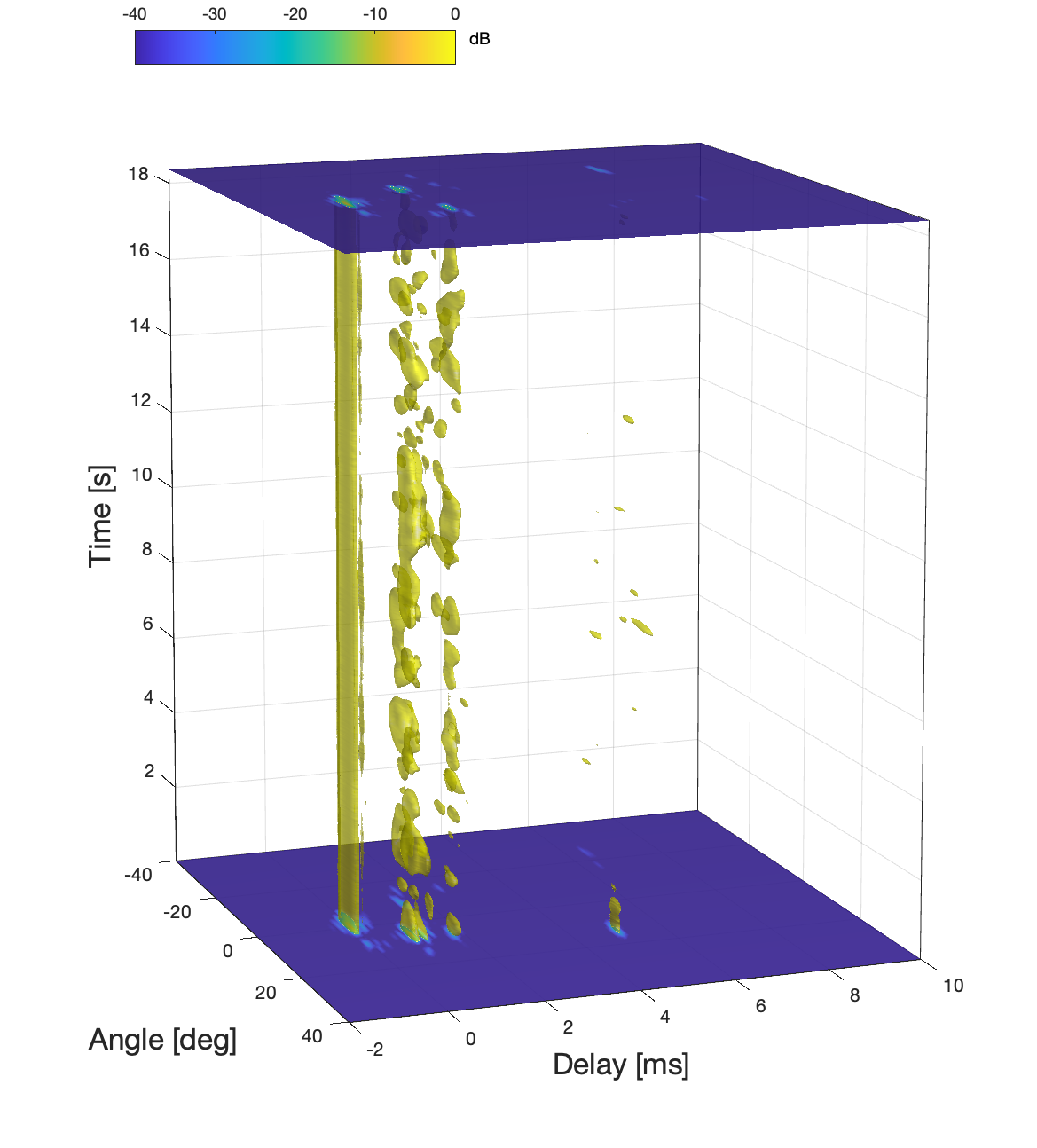}
        \caption{Delay-angle-time plot}
        \label{fig:SPACE08_average_delay_angle_time}
    \end{subfigure}
    \caption{ Average power distribution as a function of delay, angle of arrival, and time for the SPACE channel.}
\end{figure}

In this section, we present results from two at-sea experiments: the Surface Processes Acoustic Communications Experiment (SPACE) and the Mobile Acoustic Communications Experiment (MACE), conducted in 2008 and 2010, respectively. These experiments involve different channel geometries, receiver-array geometries, and signal parameters, as summarized in Table~\ref{tab:experiment_params}.
\begin{table}[H]
\centering
\caption{SPACE and MACE Signal Parameters}
\begin{center}
\begin{tabular}{c|c|c}
\hline
\hline
 & SPACE & MACE\\
\hline
center frequency $f_c$ [kHz] & 12.5 & 13\\
\hline
sampling frequency $f_s$ [kHz] & $10^7/256$ & $10^7/256$\\
\hline
symbol rate $R$ [symbols/s]  & $6510.4$ & $4882.8$\\
\hline
M-sequence length  & $2^{12}-1$ & $2^{11}-1$\\
\hline
modulation  & BPSK & BPSK \\
\hline
number of array elements ($M$)  & 24 & 12\\
\hline
element spacing $\delta$ [m]  & 0.05 & 0.12\\
\hline
\hline
\end{tabular}
\label{tab:experiment_params}
\end{center}
\end{table}
\vspace{-0.5em}
Each SPACE transmission contained $86$ repeated BPSK-modulated M-sequences of length $4095$. Fig.~\ref{fig:SPACE08_average_delay_angle} shows the channel power distribution in delay and angle. The red crosses indicate theoretical paths based on the experimental geometry \cite{blair2010beamspace}, where D, B, and BS denote the direct, bottom, and bottom-surface paths, respectively. Good agreement is observed between theory and measurements. Fig.~\ref{fig:SPACE08_average_delay_angle_time} shows stable direct and bottom paths, while the surface-reflected paths fluctuate. We consider three transmissions from Julian days 288, 294, and 300, corresponding to calm, moderate, and stormy sea states.

Each MACE transmission contained $128$ repeated BPSK-modulated M-sequences of length $2047$. The transmitter moved relative to the receiver at approximately $0.5$--$1.5$~m/s. We consider three transmissions acquired at different times on Julian day 176, representing different platform-motion patterns.
% During the SPACE experiment, each transmission included $86$ repeated BPSK-modulated M-sequences of length $4095$. Figure~\ref{fig:SPACE08_average_delay_angle} shows the channel power distribution in delay and angle. The red crosses indicate theoretical paths using the geometry of the experiment \cite{blair2010beamspace}, where D, B, and BS denote direct, bottom, and bottom-surface paths. Good agreement is observed between theory and data. Figure~\ref{fig:SPACE08_average_delay_angle_time} shows that the direct and bottom paths remain stable, whereas the surface-reflected paths fluctuate. We consider three transmissions from Julian days 288, 294, and 300, corresponding to sea states ranging from calm to stormy.

% During the MACE experiment, each transmission consisted of $128$ repeated BPSK-modulated M-sequences of length $2047$. The 12-element array, with total aperture $1.32$~m, was deployed at a depth of $40$~m, while the transmitter moved relative to the receiver at speeds between approximately $0.5$ and $1.5$~m/s. We consider three transmissions acquired at different times on Julian day 176, representing different platform-motion patterns.

We post-process the recordings by estimating the angle of the principal path at a random starting time, constructing the beamforming weights, and applying them after a delay $T_f$. For comparison, the time-reversal technique estimates the channel responses at the same instant, normalizes, conjugates, and time-reverses them before applying the filters after the same delay. We use $T_f=0.5$ s for SPACE and $T_f=3$ s for MACE. The procedure is repeated over $1000$ realizations.

% We post-process the recordings by estimating the principal-path angle using the method introduced in \cite{cuji2023bftx} at a random starting time, constructing the beamforming weights, and applying them after $T_f$ seconds to account for feedback delay and processing time. Similarly, the time-reversal method estimates the channel impulse responses at the same starting time. The estimated responses are normalized to unit energy, conjugated, time-reversed, and convolved with the recorded signals after $T_f$ seconds, after which the resulting signal is equalized. We set $T_f=0.5$~s and $T_f=3$~s for the SPACE and MACE datasets, respectively. The procedure is repeated over $1000$ realizations to account for channel and noise variability.

The receiver employs an \ac{rls}-based decision-feedback equalizer. For SPACE, the equalizer uses filter lengths of \(N_f=20\) and \(N_b=20\), while the \ac{pll} gains are \(K_{f_1}=10^{-4}\) and \(K_{f_2}=K_{f_1}/10\); the forgetting factor is \(\lambda=0.995\). For MACE, the equalizer uses filter lengths of \(N_f=15\) and \(N_b=8\), while the \ac{pll} gains are \(K_{f_1}=10^{-2}\) and \(K_{f_2}=K_{f_1}/10\); the forgetting factor is also \(\lambda=0.995\).

\begin{figure}[ht]
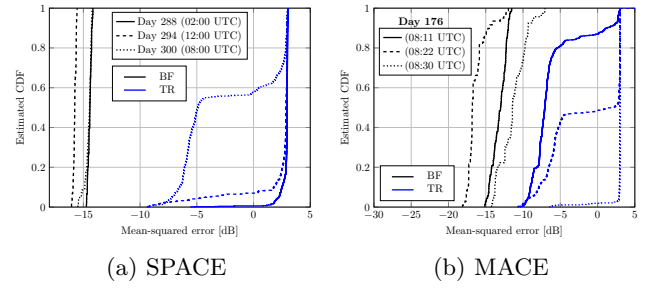

    \centering
    \begin{subfigure}[b]{0.23\textwidth}
        \scalebox{0.41}{\plotspacemsecdf}
        \caption{SPACE}
        \label{fig:space_mse_cdf}
    \end{subfigure}
    \begin{subfigure}[b]{0.23\textwidth}
        \scalebox{0.41}{\plotmacemsecdf}
        \caption{MACE}
        \label{fig:mace_mse_cdf}
    \end{subfigure}
    \caption{\Ac{cdf} of $\text{MSE}_i$ for three recorded transmissions.}
    \label{fig:space_mace_mse_cdf}
\end{figure}

Performance is evaluated in terms of the data-detection \ac{mse}
\begin{equation}
    \text{MSE}_i = \frac{1}{N_d-N_t} \sum_{n=N_t}^{N_d-1} \left|d_i[n] - \hat{d}_i[n]\right|^2,
    \label{eq:mse_super_block}
\end{equation}
where $\hat{d}_i[n]$ is the estimate of symbol $d_i[n]$ in the $i$th frame and $N_t = 4(N_f + N_b)$ is the number of training symbols. Each frame contains $N_d = 10(2^{12}-1)$ symbols for SPACE and $N_d = 10(2^{11}-1)$ for MACE. Since the channel is random, $\text{MSE}_i$ is also random; its \ac{cdf} is shown in Fig.~\ref{fig:space_mse_cdf} and Fig.~\ref{fig:mace_mse_cdf} for the SPACE and MACE datasets, respectively. Black curves (BF) denote the proposed beamforming method, while blue curves (TR) denote the time-reversal method. For SPACE, the average \ac{mse} ranges from $-15.76$ to $-14.42$~dB for beamforming and from $-4.91$ to $2.96$~dB for time reversal. For MACE, the corresponding ranges are $-16.79$ to $-11.42$~dB and $-7.08$ to $2.99$~dB. The proposed method consistently outperforms time reversal.

% Performance is evaluated in terms of the data-detection \ac{mse}
% \begin{equation}
%     \text{MSE}_i = \frac{1}{N_d-N_t} \sum_{n=N_t}^{N_d-1} \left|d_i[n] - \hat{d}_i[n]\right|^2
%     \label{eq:mse_super_block}
% \end{equation}
% where $\hat{d}_i[n]$ is the estimate of symbol $d_i[n]$ in the $i$th frame and $N_t = 4(N_f + N_b)$ is the number of training symbols. Each frame contains $N_d = 10(2^{12}-1)$ symbols for SPACE and $N_d = 10(2^{11}-1)$ for MACE. Since the channel is random, $\text{MSE}_i$ is also random; its \ac{cdf} is shown in Fig.~\ref{fig:space_mse_cdf} and Fig.~\ref{fig:mace_mse_cdf} for the SPACE and MACE datasets, respectively. The black curves labeled with ``BF" correspond to the proposed beamforming method and the blue curves labeled with ``TR" correspond to the time-reversal technique. The SPACE results yield average \ac{mse} values between $-15.76$ and $-14.42$~dB for the beamforming method and \ac{mse} values between $-4.91$ and $2.96$~dB for the time-reversal method. The MACE results yield average \ac{mse} values between $-16.79$ and $-11.42$~dB for the beamforming method and \ac{mse} values between $-7.08$ and $2.99$~dB for the time-reversal method.

% \vspace{-1em}
\subsection{Enabling Multi-User Communication}

We consider a two-user scenario by adding a simulated QPSK signal for a second user to the experimentally recorded signal. The simulated signal is upconverted to $f_c$, arrives via a single propagation path with $\theta_{0,2}=8^\circ$, and includes Doppler distortion corresponding to a user moving toward the array at $1$~m/s. The recorded and simulated signals are combined such that the ratio of the recorded signal power to the simulated signal power (signal-to-interference ratio) is $0$~dB. Zero-forcing beamformers steer toward $\theta_{0,1}=-8.7^\circ$ and $\theta_{0,2}=8^\circ$ while placing nulls toward the unintended user \cite{cuji2022multi}. Each user employs an equalizer similar to that shown in Fig.~\ref{fig:equalization}, using the parameter settings described in the previous section.

% We consider a two-user scenario based on the experimentally recorded signal, to which a simulated signal intended for a second user is added. The simulated signal consists of a QPSK-modulated sequence that is upconverted to $f_c$. Two effects are incorporated into the simulated signal: 1) it arrives via a single propagation path with an arrival angle of $\theta_{0,2} = 8^\circ$, and 2) Doppler distortion is introduced to emulate a user moving toward the array at $1$~m/s. The simulated signal is added such that the ratio of the recorded signal power to the simulated signal power (signal-to-interference ratio) is $0$~dB. The beamforming vectors are designed to steer beams toward their intended users while placing a null in the direction of the other user \cite{cuji2022multi}. Specifically, the beams for the original and second users are steered toward angles $\theta_{0,1} = -8.7^\circ$ and $\theta_{0,2} = 8^\circ$, respectively. Each user's receiver employs an equalizer similar to the one shown in Fig.~\ref{fig:equalization}, with parameter settings matching those described in the previous section.

\begin{figure}[ht]
    \centering
    \begin{subfigure}[b]{0.48\textwidth}
        \begin{subfigure}[b]{0.32\textwidth}
            \includegraphics[width=1\textwidth]{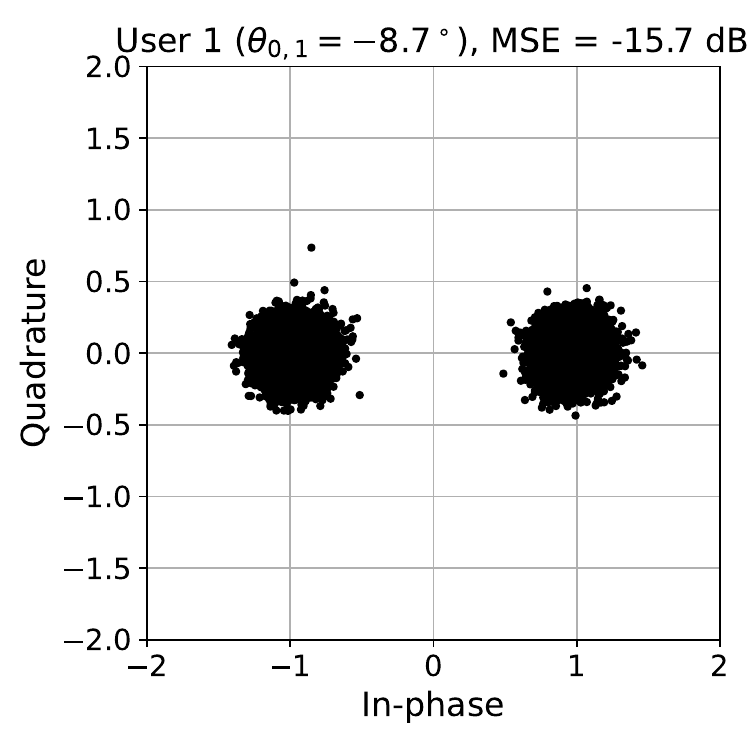}
            \caption{User 1}
        \end{subfigure}
        \begin{subfigure}[b]{0.32\textwidth}
            \includegraphics[width=1\textwidth]{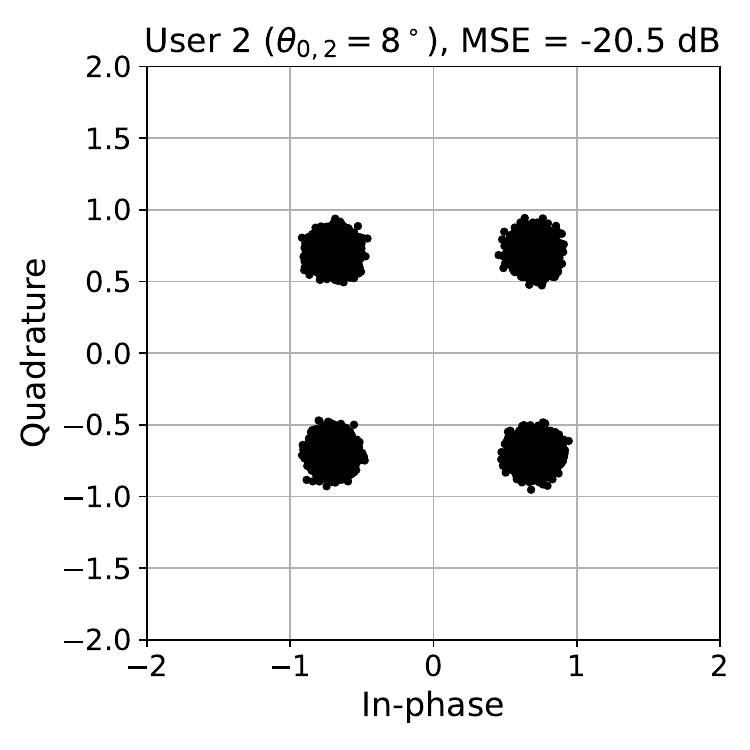}
            \caption{User 2}
        \end{subfigure}
        \begin{subfigure}[b]{0.32\textwidth}
            \includegraphics[width=1\textwidth]{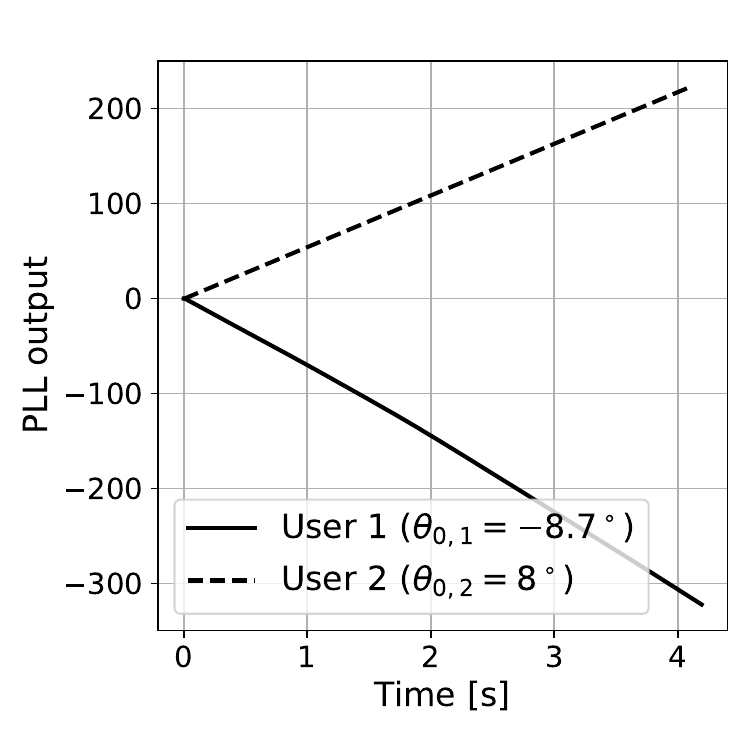}
            \caption{\ac{pll} outputs}
        \end{subfigure}
    \end{subfigure}
    \caption{Constellations of the detected data symbols and corresponding \ac{pll} outputs for both users in the asynchronous two-user transmission scenario.}
    \label{fig:mace_constellation_pll}
\end{figure}
The results are summarized in Fig.~\ref{fig:mace_constellation_pll}, which shows the constellations of the detected data symbols for both users and the corresponding \ac{pll} outputs. The \ac{mse} values of approximately $-15.7$~dB and $-20.5$~dB, together with zero bit errors, confirm reliable multi-user communication. The \ac{pll} outputs indicate effective Doppler synchronization, with the decreasing output for user~1 and increasing output for user~2 consistent with one user moving away from the array and the other moving toward it.
% \vspace{-1.0em}
\section{Conclusion \label{sec:conclusion}}
This paper investigated angle-based transmit beamforming for single-carrier underwater acoustic communications. Experimental results using the SPACE and MACE datasets demonstrated reliable single-user communication under different environmental and motion conditions, achieving low data-detection \ac{mse} and zero bit errors in the tested cases. A multi-user experiment further showed support for simultaneous asynchronous transmissions through principal-path beamforming and null steering. These results demonstrate the potential of angle-based transmit beamforming for practical underwater acoustic systems, even with long feedback delays.
% This paper investigated angle-based transmit beamforming for high-rate underwater acoustic communications. The transmitter includes a uniform linear array and the receiver implements synchronization, Doppler tracking, and adaptive equalization. Experimental results using SPACE and MACE data demonstrated reliable single-user communication under different environmental and motion conditions, delivering excellent data detection \ac{mse} and zero bit errors in the tested cases. Finally, we presented a multi-user scenario, again showing that the proposed framework can support simultaneous asynchronous transmissions through principal path beamforming and null steering. These results suggest that angle-based transmit beamforming is a promising approach for practical underwater acoustic systems and is capable of withstanding long feedback delays.

Acknowledgment: This work was supported by the Office of Naval Research under Grant \texttt{N00014-23-1-2852} and Stony Brook University.
% \section{Acknowledgment}
% This work was supported by the Office of Naval Research under Grant \texttt{N00014-23-1-2852} and Stony Brook University.

% \input{tex/abstract}
% \input{tex/intro}
% \input{tex/signal}
% \input{tex/bf}
% \input{tex/experiment}
% \input{tex/conclusion}

% \clearpage
\IEEEtriggeratref{12}
\bibliographystyle{IEEEtran}
\bibliography{ref}

\end{document}